\documentclass[amsmath,amssymb, amsfonts]{article}

\usepackage{epsf}
\usepackage{rotating}
\usepackage{graphicx,epsfig}
\usepackage{dcolumn}
\usepackage{bm}

\begin{document}

\author{C. H. Yeung$^{1, 2}$ and Y.-C.~Zhang$^{2}$}
\title{Minority Games}
\maketitle

\begin{quote}
$^{1}$ Department of Physics,
The Hong Kong University of Science and Technology, Hong Kong, China\\
$^{2}$ D\'epartement de Physique,
Universit\'e de Fribourg,\\
P\'erolles, Fribourg, CH-1700 Switzerland
\end{quote}



\tableofcontents

\section{Glossary}
$\\$
\noindent {\it Cumulated payoff}
$\\$

It refers to the reward accumulation or the score counting 
for all the individual strategies being held by the agents in the game.
Each strategy has its own value of cumulated payoff.
When a strategy gives a winning or losing prediction for the next round of the game
(no matter whether the agents follow this prediction), 
scores are added to or deducted from this strategy respectively.
It is also known as virtual point or virtual score of the strategies.
The way of rewarding or penalizing is known as the payoff function.

$\\$
\noindent {\it Attendance}
$\\$

In the contexts of the Minority Game,
attendance refers to the collective sum of all agents' actions at each round of the game.
For the ordinary games, 
it is equal to the difference in the number of agents in choosing the two different choices or actions
(in early formulation, 
it is equal to the number of agents in choosing one of the particular choice).
The terminology ``attendance" origins from the ancestor form of the Minority Game,
the ${\it El Farol}$ Bar problem by W. B. Arthur,
in which agents choose whether to attend a bar at every round of the game.

$\\$
\noindent {\it Volatility}
$\\$

Volatility in Minority Games is the time average variance of the attendance after each 
round of the game.
It is an inverse measure of the efficiency of resource distribution in the game.
A high volatility corresponds to large fluctuations in attendance and 
hence an inefficient game.
A low volatility corresponds to smaller fluctuations in attendance and
hence an efficient game.

$\\$
\noindent {\it Predictability}
$\\$

It is an important macroscopic measurable in the game
and also the order parameter which characterizes the major phase transition in the system.
It is usually denoted by $H$ which
is a measure of non-uniform probabilistic outcome 
of the attendance given a certain information provided to the agents.
A higher predictability refers to the case where attendance tends to be 
positive or negative for a certain piece of information,
which makes the game outcome more predictable.

$\\$
\noindent {\it Symmetric and asymmetric phase}
$\\$

The two important phases of the system.
The symmetric phase is also known as crowded phase or the 
unpredictable phase.
The asymmetric phase is also known as uncrowded phase, 
the dilute phase or the
predictable phase.
The system's behaviours, 
dynamics and characteristics are different in this two phases.
The two phases are characterized by the order parameter,
called the predictability $H$.

$\\$
\noindent {\it Endogenous and exogenous games}
$\\$

Endogenous games refer to games which utilize the past winning history 
to generate signals or informations for agents to make decisions 
in the next round. 
Endogenous games are also known as games with real history.
Exogenous games refer to games which utilize random signals or 
informations for agents to make decisions
in the next round.
Exogenous games are also known as games with random history,
or external information.

$\\$
\noindent {\it On-line Update and Batch Update}
$\\$

On-line update refers to the evaluation of payoffs of strategies after
each round of the game.
Thus, 
the priority of strategies being employed by an agent may be altered after any round of the game.
Exogenous game or random history game sometimes employ the batch update method
which refers to the evaluation of payoffs on strategies only after a fixed number of rounds where
all the possible signals or informations have been appeared. 
In games with ordinary batch update,
all the possible signals appear once in each batch before the update of payoffs on strategies,
the order of appearance of signals in each batch is thus irrelevant.

\section{Definition of the Subject and Its Importance}

The Minority Game (abbreviated, MG) refers to the simple adaptive 
multi-agent model of financial markets with the original formulation 
introduced by Challet and Zhang in 1997. 
In this model of repeated games,
agents choose between one of the two decisions at each round of the game,
using their own simple inductive strategies.
At each round, 
the minority group of agents win the game
and rewards are given to those strategies that predict the winning side.
Daily examples of minority game include drivers choosing a less crowded road
or people choosing a less crowded restaurant.
Unlike most economics models or theories that assume investors are deductive in nature,
a trial-and-error inductive thinking approach 
is implicitly implemented in process of decision making when agents choose their choices in the games.
In this original formulation,
the history or the information given to agents is a string of binary bit which is composed of the winning sides
in the past few rounds.

While the original model is simple,
many variants of the models came out after the original model.
In some other context and later literatures,
the term ``Minority Games" is sometimes referred to as a class of multi-agent models
which contains all the variants of the original Minority Game.
Most of the models in this class of game 
share the principal features that agents are inductive in nature.
Thus,
strategies with accumulated virtual score are usually present in this set of models.
As a result, 
the original formulation of the Minority Game by Challet and Zhang in 1997 
is sometimes referred as the ``original Minority Game" or the ``basic Minority Game".

With attempts of investigating economical dynamics,
physicists found most of the economics models deductive in nature.
Since investors have expectation of the future,
economical models conceptually differ from conventional physical models in which variables are 
only history dependence.
As a result,
it becomes difficult for physicists to develop and analyze the traditional financial models,
even with well-developed mathematical tools.
The {\it El Farol} bar problem and the Minority Game somehow tackle the problem
by assuming investors can be inductive in nature with bounded rationality,
which they predict the future by only examining the past states of the system,
similar to the ordinary physics models.

In the physics community,
the basic Minority Game and its variants are an interesting and newly established 
class of complex and disordered systems which contain a large amount of physical aspects.
In addition to the modeling purpose of the financial markets,
it is also a simple model
where Hamiltonian can be defined and analytic solutions are developed in some regime of the model,
from which the model is viewed with a complete physical sense.
It is also characterized by a clear two-phase structure
with very different collective behaviours in the two phases,
as in conventional physical systems.
All these physical properties further raise the interests of 
physicists in understanding and solving the model analytically,
using the techniques origin from statistical mechanics.
Other than these collective behaviours,
physicists are also interested in the dynamics of the games.
Periodic attractors, anti-persistence  and crowd-anticrowd movement of agents
are also observed.
In this way, 
the Minority Game and its variants serve as a useful tool and provide a new direction for physicists
in viewing and analyzing the underlying dynamics in the financial markets,
and at the same time analytical techniques from statistical physics can be widely applicable.

On the other hand, 
for modeling purposes,
Minority Games serve as a class of simple models 
which are able to produce some of macroscopic features
being observed in the real financial markets.
Such features are usually termed as stylized facts which 
include the fat-tail price return distribution and volatility clustering.
Crashes and bubbles are also observed in some of the variants and other models
inspired by the Minority Game.
The grand-canonical versions of the game suggest the conjecture of financial markets
being a critical phenomenon in physics.

Due to the simplicity of the original model,
a large freedom is found in modifying the models to make the models more realistic
and closer to the real financial markets.
Many details in the model can be fine-tuned to imitate the real markets.
Minority Games setup a framework of agent-based models where predictability of financial data may be possible.
Sophisticated models based on games can be setup and implemented on real trading,
which show a great potential over the commonly adopted statistical techniques in financial analysis,
As a result,
Minority Games raise the interests of some of the economists in switching to
employ agent-based models 
in understanding the underlying mechanism behind the socio-economics systems.
Minority Games also shift the emphasis of some economists in
investigating the formation of price pattern, 
rather than just data analysis of the price pattern.

\section{Introduction}

The basic Minority Game is formulated by Damien Challet and Yi-Cheng Zhang in 
1997 \cite{challet97} with their work being published in a statistical physics journal.
This model is inspired by the {\it El Farol } Bar problem introduced by W. Brian Arthur in 1994 \cite{arthur94}
with his work being presented and published in an economical meeting and its proceedings.
This already shows the interdisciplinary nature of the Minority Game with an economical origin,
in a physical perspective.
The Minority Game follows the major conceptual structure being implemented 
in the {\it El Farol} Bar problem,
with some modifications on the model structure.

In the original {\it El Farol} Bar problem, 
each individual of a population choose whether to attend a bar on every thursday evening.
The bar has limited number of seats and can at most entertain $x\%$ of the population.
If less than $x\%$ of the population go to the bar,
the show in the bar is considered to be enjoyable and it is better to attend the bar rather than 
staying at home.
On the other hand,
if more than $x\%$ of the population go to the bar,
all the people in the bar would have an unenjoyable show and staying at home
is considered to be better choice than attending the bar.
In order to make decisions on whether to attend the bar,
all the individuals are equipped with certain number of strategies.
These strategies provide them the predictions of the 
attendance in the bar next week,
based on the attendance in the past few weeks.
All individuals rank their strategies according to their past performance
and make decisions by considering the attendance predicted by their own best strategy.

Several changes are made in the model when the Minority Game 
is formulated from the {\it El Farol} Bar problem.
Instead of using the history of past attendance,
a string of binary bits which records the past few winning predictions or actions
are employed as information.
The predictions of the strategies are the winning choices in the next round,
with no prediction about the actual size of attendance.
Thus,
binary information and predictions are implemented,
which greatly reduce the dimensional space of the system.
In addition,
the winning choice is determined by the minority choice (instead of the parameter $x$ in the Bar problem)
at every round of the game,
hence the two choices are symmetric.
Due to the minority rule,
the population is restricted to be an odd integer in the original formulation.

These modifications of binary and symmetric actions make the model more accessible
for the physics community.
The first publication of the Minority Game raised a great interest for 
some of the statistical physicists to carry on researching the Minority Game 
and formulate its variants.
Some of the physicists begin to identify the study of such class of models
as in the field of econophysics.
In 1999, 
R. Savit {\it et al.} \cite{savit99} published their work on the analysis of the Minority Game,
which is crucial to subsequent theoretical developments of the game.
They discovered an important control parameter $\alpha$, 
which is defined as the ratio of total amount of possible information to the population size.
It rescales the macroscopic observables of the game for
different amount of information and population size.
A phase transition is observed at the critical value of $\alpha$ which separate the two phases,
namely the symmetric phase and the asymmetric phase.

After their discovery on the rescaling properties and the phase transition of the Minority Game,
great efforts were put to solve the model analytically,
using well-developed techniques in the field of statistical physics
\cite{johnson99b, hart01, hart00, challet00d, challet00, marsili00, marsili01b, challet99,
garrahan00, sherrington02, heimel01, heimel01a, coolen01, deMartino01}.
In order to solve the model,
the basic Minority Game is sometimes modified to increase the feasibility of analytical approach.
In some variants of the Minority Game,
the model is simplified to preserve only the major dynamical behaviours while
in some other variants,
features are added to the game which make the model more
comparable to traditional physical models.
As a result,
a large number of variants of the Minority Game come out
with the attempt of analytic description.

On the other hand,
attempts are also made to make the models more comparable to real financial markets.
Some physicists and even economists modify the basic model by 
putting in more features from the real markets
\cite{marsili01, andersen03, johnson99, challet08, yeung08, slanina99, jefferies01, challet00c, challet03, challet01, giardina03}.
Stylized facts are found in the critical regime of the grand-canonical version 
\cite{slanina99, jefferies01, challet00c, challet03, challet01, giardina03}
of the Minority Game
in which agents can choose to refrain from participation in the game.
This suggests the conjecture of financial markets being in critical state
and further push the development of the model towards this direction.
Some new models are also developed to include 
more financial aspects.
Efforts are made to have a better understanding of the market 
through the agent-based approach.
The macroscopic observations from the models become 
more realistic but at the same time,
the models become more sophisticated.
Due to the analytic goal in solving the model in a physical sense and 
for the modeling purposes,
there are vast number of variants of the Minority Game which 
make it a class of models.

In the following sections,
we briefly describe the formulation of the basic model and its variants,
and briefly introduce the physical properties, 
the analytic approaches of the model and its link with financial markets.
We review the formulation of the basic Minority Game in Section \ref{secMG}.
Some major physical properties of the basic Minority Game are given in Section \ref{secPhys},
the effect of temperature is also discussed which 
was originally introduced in the Thermal Minority Game (TMG).
In Section \ref{secVariants},
we review briefly some important variants of the Minority Game and their physical significance, 
these include the Evolutionary Minority Game (EMG), 
the TMG,
the Minority Game without information and the Grand-canonical Minority Game (GCMG)
while their corresponding implications for the financial markets 
and some other variants will be discussed in Section \ref{secFin}.
In Section \ref{secAnalytic},
we briefly introduce some of the analytic approaches on the Minority Game.
In Section \ref{secFin},
we review some of the financial features produced by Minority Games
and their implications.
Finally in Section \ref{secFutureDirection},
we describe some of the possible directions for the future development of the Minority Game.

\section{The Minority Game}
\label{secMG}

The basic Minority Game \cite{challet97} is defined as follows.
We consider a population of $N$ agents competing in repeated games,
where $N$ is an odd integer.
At each round of the game, 
each agent has to choose between one of the two actions, 
namely ``0" and ``1"
(in most of the subsequent literatures, 
``-1" and ``1" instead of ``0" and ``1" are used as the actions,
we shall keep the following discussions using the actions ``-1" and ``1"),
which can also be interpreted as the``sell" and ``buy" actions.
These actions are sometimes called the bid and is denoted by $a_i(t)$, 
corresponding to the bid of agent $i$ at time $t$.
The minority choices win the game at that round and all the winning agents are rewarded.

\begin{table}
\centering
\begin{tabular}{|cc|}
\hline
History & Prediction\\
\hline
000 & 1 \\
001 & 0 \\
010 & 0 \\
100 & 1 \\
011 & 0 \\
101 & 1 \\
110 & 0 \\ 
111 & 0 \\
\hline
\end{tabular}
\caption{An example of strategy with $M=3$}
\label{strTable}
\end{table} 

Before the game starts, 
every agent draws $S$ strategies from a strategy pool which help them to make decisions throughout the game.
There is no {\it a priori} best strategy.
These strategies can be visualized in the form of tables where each strategy 
contains a ``history column" (or ``signal" column) and a ``prediction column".
Each row of the history column is a string of $M$ bits,
which represents the {\it history} of the past winning actions in the previous $M$ steps,
which is also known as {\it signal} or {\it information}.
The history is evolving with time and is usually denoted by $\mu(t)$.
The parameter $M$ is sometimes known as the {\it brain size} or the {\it memory} of the agents.
An example of strategy with $M=3$ is given in Table~\ref{strTable}.
For games with memory $M$,
the total number of possible signals is $2^M$
and thus the total number of possible strategies in the strategy pool is $2^{2^M}$.
We note that even a relatively small $M$,
such as $M=5$,
the total number of possible strategies is already huge.

As shown in the strategy in Table \ref{strTable}, 
a history of ``110" corresponds to the case where the past 3 winning actions are 
``1", ``1" and ``0",
and the corresponding prediction of winning choice for the next round is ``0".
Strategies can be conveniently represented by $P$-dimensional vectors which record only the $P$ predictions,
where $P=2^M$.
If the strategy gives a correct prediction on the winning choice,
one point is awarded to the strategy.
All the $S$ strategies of an agent have to predict at every round of the game,
and points are given to those strategies (no matter whether they are
being selected by the agent to make real actions)
that give correct predictions.
The scores of all the strategies are accumulated which 
are thus known as the {\it virtual points},
{\it virtual scores} or the {\it cumulated payoffs} of the strategies.
These scores start at zero in the basic Minority Game.
At every round of the game,
agents make their decisions according to the strategy with the highest virtual score
at that particular moment.
If there are more than one strategies with the highest score,
one of these strategies is randomly employed.
Agents themselves who make the winning decisions are also
rewarded with points,
and is called the {\it real points} of the agents
(to be distinguished from the {\it virtual points} of the strategies).

More explicitly,
we define the {\it attendance} $A(t)$ as the collective sum of actions from all agents at time $t$.
If we denote the prediction of strategy $s$ of agent $i$ under the information $\mu(t)$
to be $a_{i,s}^{\mu(t)}$ at time $t$,
which can be either ``-1" or ``1" ,
each strategy can be represented by a $P$-dimensional vector $\vec{a}_{i,s}$ where
all the entries are either ``-1" or ``1".
The attendance $A(t)$ can then be expressed as 
\begin{eqnarray}
\label{attendance}
	A(t) = \sum_{i=1}^N a_{i,s_i(t)}^{\mu(t)} =\sum_{i=1}^N a_i(t)
\end{eqnarray}
where $s_i(t)$ denotes the best strategy of agent $i$ at time $t$,
i.e .
\begin{eqnarray}
\label{bestStr}
	s_i(t) = \arg \max_s U_{i,s}(t)
\end{eqnarray}
and $a_i(t)$ denotes the the real actions or so-called the {\it bids} of the agents,
i.e.
\begin{eqnarray}
\label{realBid}
	a_i(t) = a_{i,s_i(t)}^{\mu(t)}
\end{eqnarray}
With this $A(t)$, 
the cumulative virtual score or payoff $U_{i,s}$ 
of the strategy $s$ of agent $i$ can be updated by 
\begin{eqnarray}
	U_{i,s}(t+1) = U_{i,s}(t)-{\rm sign}[a_{i,s}^{\mu(t)}A(t)]
\label{stepPayoff}
\end{eqnarray}
where ${\rm sign}(x)$ is the sign function
(in some literatures where ``0" and ``1" are employed as actions,
the last term in Eq.~(\ref{stepPayoff}) becomes $-{\rm sign}[(2a_{i,s}^{\mu(t)}-1)A(t)])$
Here one point is added or deducted from the strategies which 
give a correct or wrong prediction respectively,
and is usually called the {\it step payoff} scheme.
We note that the negative sign in Eq.~(\ref{stepPayoff}) corresponds to 
the minority nature of the game,
i.e. when $a_{i,s}^{\mu(t)}$ and $A(t)$ are of opposite signs,
a point is added to the strategy.
The real gain of agent $i$ at time $t$ is $-{\rm sign}[a_i(t)A(t)]$.

Thus,
every agent is considered to be adaptive,
who can choose between their $s$ strategies and the 
relative preference of using strategies is changing with time 
and adaptive to the market outcomes.
They are also considered to be inductive,
who base their decisions according to the best choice they know,
with their limited number of strategies,
but not the global best choice given by all possible strategies
(the one with highest virtual score among the entire strategy space).
The game is also a self-contained model,
in which the agents make individual actions according to the history,
individual actions are summed up to give the history for the next round
which is then used by the agents to make predictions again.

As the total number of agents $N$ in the game is an odd integer,
the minority side can always be determined 
and the number of winners is always less than the number of losers,
implying the Minority Game to be a {\it negative sum} game.
Due to the minority nature of the game and as the two actions are symmetric,
the time average of $A(t)$ always has a value of $0$ 
(or $\langle A(t)\rangle$ = $N/2$ if ``0" and ``1" are employed as actions).
Hence,
instead of the average attendance, 
one may be more interested in the fluctuations of attendance around the average values
and this turns out to be the one of the most important macroscopic observables in the subsequent development.
We denote $\sigma^2$ to be the variance of attendance, 
or also known as the {\it volatility},
given by
\begin{eqnarray}
\label{variance}
	\sigma^2=\langle A^2\rangle-\langle A\rangle^2
\end{eqnarray}
with $\langle A\rangle=0$ for games with actions ``-1" and ``1".
$\sigma^2$ is an inverse measure of the market efficiency in the game.
We consider two extreme cases of game outcomes.
For the first one,
there is only one agent choosing one side while all the others choose the opposite side.
There is a single winner and $N-1$ losers which
is considered to be highly inefficient in the sense of resource allocation,
and the supply and demand are highly unbalanced.
For the second case,
$(N-1)/2$ of the agents choose one side while $(N+1)/2$ of the agents choose the opposite side.
There are $(N-1)/2$ winners and the supply and demand is maximally balanced.
Thus,
one may expect to minimize fluctuations of attendance 
for advantages of agents as a whole.

Some simplifications or modifications to the basic Minority Game 
are suggested and employed in later literatures,
where the major physical features of the models are preserved.
A. Cavagna \cite{cavagna99b, cavagna00b} observed that the variance of attendance is almost unaffected
if the history string is replaced by a random {\it invented} string,
provided that all agents receive the same string at the same time.
That is,
instead of their self-generated winning history,
they react to a virtual random information which is completely unrelated to the previous winning groups.
In this case,
the signal strings are usually called {\it information} instead of {\it history}.
The games which feedback the real history are known as the {\it endogenous} games
while those games which employ random history is called the {\it exogenous} games.
In exogenous games,
the total number of signal is no longer restricted to be $2^M$,
instead it can be any integer which is usually denoted by $P$, 
and is sometimes known as the complexity of information.
Every random signal or information appears with a probability of $1/P$.
These random informations make the dynamics of the game more stochastic,
which is an extreme advantage for analytic approaches.
In endogenous game,
$P=2^M$.

As ``-1" and ``1" are employed as actions,
instead of adding or deducting one virtual point to the strategies,
the cumulated payoff $U_{i,s}$ can be updated by the following equation with {\it linear payoff} scheme
\begin{eqnarray}
\label{payoff}
	U_{i,s}(t+1) = U_{i,s}(t) - a_{i,s}^{\mu(t)}A(t)
\end{eqnarray}
where $A(t)$ is the attendance given by Eq.~(\ref{attendance}).
A factor of  $1/N$, 
$1/\sqrt{N}$ or $1/P$ are always employed to rescale the last term.
While the real gain for agent $i$ is $-a_i(t)A(t)$,
the total gain for all the agents is $\sum_i -a_i(t) A(t)=-A^2(t)$,
preserving the negative sum nature of the game.
This modification is important for analysis 
while the qualitative behaviours of the game are preserved.
It also has the meaning of having a higher reward or larger penalty if
a smaller minority or a larger majority group is predicted respectively.

In the original game,
the ordinary strategy space has a size of $2^{2^M}$.
Challet {\it et al.} \cite{challet98} showed that a {\it reduced strategy space} (RSS) can be employed
in which the qualitative behaviours of the game and 
the numerical values of variance are not largely affected.
We first construct a set of $2^M$ uncorrelated strategies,
in which every two strategies of the set have exactly half of the predictions different.
The reduced strategy space is then formed by combining this set with 
the set of their anti-correlated strategies,
in which every strategies in the latter set have exactly opposite predictions
to their anti-partners in the former set.
Hence,
the size of the reduced strategy space is $2^M\cdot 2=2^{M+1}$.
This virtue of the reduced strategy space simplify the theoretical analysis
in crowd-anticrowd approaches \cite{johnson99b, hart00, hart01}.
Although the dimension of reduced strategy space is highly reduced,
the ordinary strategy space is commonly employed in numerical simulations.

\section{The Physical Properties of the Minority Game}
\label{secPhys}

There are several parameters being introduced in the basic Minority Game,
include $N$, $M$ (or $P$) and $S$ corresponding to the population size,
the memory of the agents (the complexity or the total amount of possible information)
and the number of strategies that each agent holds.
The predictions $a_{i,s}^{\mu}$ of strategies are fixed for every agents throughout the game
and are considered to be quenched disorders of the system in physics.
The cumulated payoffs of strategies 
which evolve with time are considered to be dynamic variables or annealed variables of the system.
The game is also a highly frustrated model.
Due to the minority nature of the model,
frustration results in the fact that not all the agents can be satisfied simultaneously.
We focus our discussions on the case of $S=2$,
where cases of larger $S$ (not extensively large) will be briefly discussed 
and have been shown in to share very similar behaviours as the case of $S=2$\cite{challet98}.

\subsection{Major features: Phase transition, Volatility and Predictability}

\begin{figure}
\centerline{\epsfig{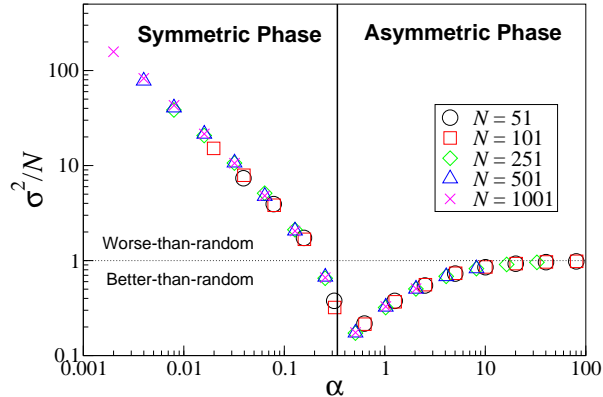}}
\caption{
The simulation results of the volatility $\sigma^2/N$ as a function of the control parameter $\alpha=2^M/N$
for games with $S=2$ strategies for each agent averaged over 100 samples.
Endogenous information and linear payoff are adopted in these simulations.
Dotted line shows the value of volatility in random choice limit.
Solid line shows the critical value of $\alpha=\alpha_c\approx 0.3374$.
Resolution of the curve can be improved to shows $\sigma^2/N$ attains minimum at $\alpha\approx\alpha_c$
}
\label{varVsAlpha}
\end{figure}

In 1999,
Robert Savit, 
Radu Manuca and Rick Riolo \cite{savit99} found that the macroscopic behaviour of the system
does not depend independently on the parameters $N$ and $M$,
but instead depends on the ratio
\begin{eqnarray}
\label{alpha}
	\alpha \equiv \frac{2^M}{N} = \frac{P}{N}
\end{eqnarray}
(which is denoted by $z$ in their original paper) which serves as the most important 
control parameter in the game.
This scaling is also true for $P \neq 2^M$ in exogenous games.
The volatility $\sigma^2/N$ and the predictability $H/N$ (which we are going to define later) 
for different values of $N$ and $M$ 
depend only on the ratio $\alpha$.
A plot of $\sigma^2/N$ against the control parameter $\alpha$
for endogenous game is shown in Fig.~\ref{varVsAlpha}.
We can see that the graph shows a data collapse of $\sigma^2/N$ for different
values of $N$ and $M$.
The dotted line in Fig.~\ref{varVsAlpha} corresponds to the coin-toss limit (random choice limit),
in which agents play by making random decisions at every rounds of the game.
This value of volatility in coin-toss limit can be obtained by simply assuming a binomial distribution of agents' actions,
with probability $0.5$,
where $\sigma^2/N=0.5(1-0.5)\cdot 4=1$.
When $\alpha$ is small,
the volatility of the game is larger than the coin-toss limit which implies the collective
behaviours of agents are worse than the random choices.
In early literatures,
it is known as the {\it worse-than-random} regime.
When $\alpha$ increases,
the volatility decreases and enter a region where agents are performing better than the random choices,
which is known as the {\it better-than-random} regime.
The volatility reaching a minimum value which is substantially smaller than the coin-toss limit.
When $\alpha$ further increases,
the volatility increases again and approaches the coin-toss limit.

These results also allow us to identify two phases in the Minority Game,
as separated by the minimum of volatility in the graph.
The value of $\alpha$ where the rescaled volatility attends its minimum 
is denoted by $\alpha_c$,
which represents the phase transition point.
$\alpha_c$ has a value of $0.3374\dots$ (for $S=2$) by analytical calculations \cite{challet00, heimel01}.
Generally,
for $\alpha<\alpha_c$,
the volatility $\sigma^2$ and the spread of volatility for different samples of simulation
are proportional to $N^2$.
Beyond the transition point for $\alpha>\alpha_c$,
the volatility $\sigma^2$ and the spread of volatility are generally
proportional to $N$.
These can be recognized by the asymptotic behaviour of the graph in Fig~\ref{varVsAlpha}
where the slope approaches $-1$ for $\alpha<\alpha_c$ and approaches $0$ 
for $\alpha>\alpha_c$.

\begin{figure}
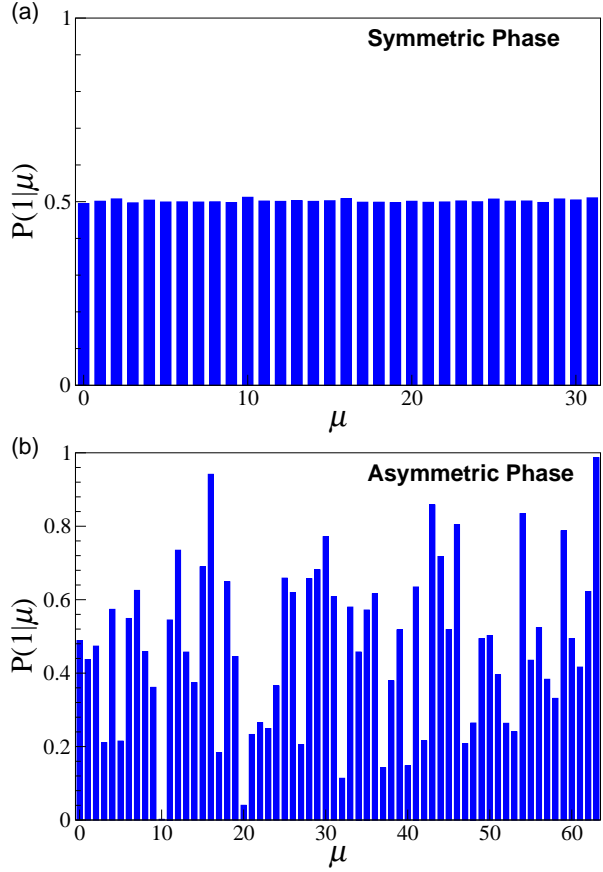

\centerline{\epsfig{figure=m5.eps, width=0.65\linewidth}}
\centerline{\epsfig{figure=m6.eps, width=0.65\linewidth}}
\caption{
The histogram of the probabilities ${\rm P}(1|\mu)$ of winning action to be ``1'' given 
information $\mu$ (plotted as the decimal representations of the binary strings of information),
for games of $N=101$ agents and $S=2$ in (a) symmetric phase with $M=5$, i.e. $\alpha\approx 0.316<\alpha_c$
and (b) asymmetric phase with $M=6$, i.e. $\alpha\approx 0.634>\alpha_c$.
Endogenous information and linear payoff are adopted.
The histogram in (a) would be even more uniform if step payoff is adopted,
as shown in the original paper \cite{savit99}.
}
\label{histogram}
\end{figure}

In addition to the different scaling of volatility with $N$,
other quantities also show different behaviours in the two phases.
By examining the distributions of winning probabilities for a particular action
after different history strings,
R. Savit {\it et al.} \cite{savit99} found that these distributions 
are completely different in the two phases.
By defining ${\rm P}(1|\mu)$,
to be the conditional probability of action ``1" turns out to be the minority group
after the history or information $\mu$, 
the histogram for ${\rm P}(1|\mu)$ is flat at 0.5 for all $\mu$ when $\alpha<\alpha_c$, 
as shown in Fig.~\ref{histogram}(a).
For $\alpha>\alpha_c$  as shown in Fig.~\ref{histogram}(b),
this histogram for ${\rm P}(1|\mu)$ is not flat and uniform.
This result is highly important which implies that below $\alpha_c$,
there is no extractable information from the history string of length $M$,
since the two actions have equal probability of winning  (both are 0.5) for any history string.
However,
beyond the phase transition when $\alpha>\alpha_c$,
there is an unequal winning probability of the two actions,
by just looking at the past $M$ winning actions of the game.
Hence,
we can call the phase for $\alpha<\alpha_c$ the {\it unpredictable} or the {\it symmetric} phase,
as agents cannot predict the winning actions from 
the past $M$-bit history (the winning probabilities are symmetric).
On the contrary,
the phase of $\alpha>\alpha_c$ is called the {\it predictable} or the {\it asymmetric} phase,
as there is bias of winning actions given the past $M$-bit history string
(the winning probabilities are asymmetric).

\begin{figure}
\centerline{\epsfig{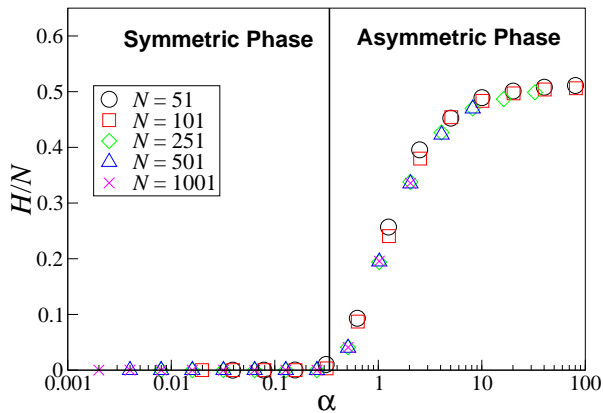}}
\caption{
The simulation results of the predictability $H$ as a function of the control parameter $\alpha=2^M/N$
for games with $S=2$ strategies for each agent averaged over 100 samples.
Endogenous information and linear payoff are adopted in these simulations.
}
\label{figH}
\end{figure}

Owing to the results in these histograms,
a useful quantity can be defined to measure the ``non-uniformity" 
of the winning probabilities or the information content given by the past $M$-bit history string.
We denote $H$ to be the {\it predictability} of the game
which is given by the following formula,
\begin{eqnarray}
\label{predictability}
	H=\frac{1}{P}\sum_{\mu=1}^{P}\langle A|\mu\rangle^2
\end{eqnarray}
with $P=2^M$ again.
$H/N$ is plotted as a function of $\alpha$ in Fig. \ref{figH}.
In the symmetric phase,
$\langle A|\mu\rangle=0$ for all $\mu$ as the actions of
``-1" and ``1" are equally likely to appear after $\mu$.
Hence, $H=0$ in the symmetric phase.
In the asymmetric phase,
$\langle A|\mu\rangle\neq 0$ for all $\mu$ as the actions of
``-1" and ``1" are not equally likely after $\mu$.
Hence, $H>0$ in the asymmetric phase.
$H$ begins to increase at $\alpha=\alpha_c$ as shown in Fig. \ref{figH}.
Analytic approaches are developed which are based
the minimization of predictability $H$ \cite{challet00d, challet00, marsili00, marsili01b}
In addition to the predictability $H$,
the fraction of {\it frozen} agents also increases drastically before $\alpha_c$ and decreases afterward.
Frozen agents are agents who always use the same strategy for making decisions.
In contrast,
{\it fickle} agents are those who always switch strategies.

In exogenous games, 
the phase transition, 
the scaling by $\alpha$ and the properties of the two phases preserve.
The symmetric and asymmetric winning probabilities also preserve
with $\mu$ representing the 
random information given to the agents in the conditional probabilities ${\rm P}(1|\mu)$
(not the actual past history) \cite{cavagna00b, savit00}.
The numerical values of volatility in the asymmetric phase has
a small deviation from that of the endogenous games.
It is because the winning probabilities are asymmetric,
the probability of history appearance is non-uniform in endogenous games,
while in exogenous games,
we assume a uniform appearing probability of all the random information $\mu$ being
given to the agents.

\subsection{Formation of Crowds and Anitcrowds and Anti-persistence in the Symmetric Phase}

In the symmetric phase with small $\alpha$,
the amount of available information $P$ is small when compared to the number of agents $N$.
Agents are able to exploit the information well and they react like a crowd which 
result in a large volatility.
From the point of view of the strategy space,
the number of independent strategies \cite{challet98, johnson99b, hart00, hart01} (as discussed in RSS) is smaller
than $N$ in the symmetric phase.
Many agents use identical strategies and react in the same or similar ways, 
forming {\it crowds} and {\it anticrowds} giving large volatility.
This is sometimes known as the herd effect in the minority game.
On the other hand,
in the asymmetric phase with large $\alpha$,
the available information $P$ is too much and complex when compared to $N$.
Agents are not able to exploit all the information 
and they react like making random decisions,
resulting in a volatility approaching the coin-toss limit.
In this case,
the number of independent strategies is greater than $N$ and 
agents are unlikely to use the same strategies, 
they acts independently and crowds are not formed.
The phase transition occurs when $\alpha_c$ is roughly $O(1)$,
where $N$ is roughly the same size as the available information $P$.

In addition to formation of crowds,
anti-persistence of the winning actions exists in the symmetric phase.
For small $\alpha$ in the symmetric phase,
consecutive occurrence of the same signal lead to opposite winning actions \cite{savit99, challet99, wong05},
which is known as anti-persistence of the Minority Game.
For example,
in the case of $M=2$,
if the history ``01" leads to a winning choice of ``0",
then the next appearance of the history ``01" will lead to a winning choice of ``1".
It results in a periodic dynamics of the game for $\alpha<\alpha_c$
with a period of $2^M\cdot 2$,
where every history appear exactly twice with different winning actions 
for the first and second occurrence.
This kind of anti-persistence disappears in the asymmetric phase.
Instead, 
persistence is more likely \cite{challet99},
in which the consecutive occurrence of the same signal tends to have 
a higher probability in giving out the same winning actions.

\subsection{Dependence on Temperature and Initial Conditions}

In 1999,
Cavagna {\it et al.} \cite{cavagna99} introduced the probabilistic fashion,
the {\it temperature},
to the decision making process of agents in the model known as Thermal Minority Game (TMG).
This stochasticity of temperature can also be implemented in the basic game.
Instead of choosing the best strategy for sure,
agents employ their strategy $s$ with the probabilities $\pi_{i,s}$ given by 
\begin{eqnarray}
\label{strProb}
	{\rm Prob}\{s_i(t) = s\} =\pi_{i,s} =\frac{e^{\Gamma U_{i,s}(t)}}{\sum_{s'}e^{\Gamma U_{i,s'}(t)}}
\end{eqnarray}
where $s_i(t)$ denote the strategy being employed by agent $i$ at time $t$.
$\Gamma$ is denoted by $\beta$ in the original formulation,
which corresponds to the {\it inverse temperature} (as in physical systems) of individual agents.
It can also be interpreted as the {\it learning rate} of the system \cite{marsili01b}.
Roughly speaking,
this is because the dynamics of scores take a time of approximately $1/\Gamma$ 
to learn a difference in the cumulated payoffs of the strategies.
For small $\Gamma$,
the system take a convergence time of order $N/\Gamma$ to reach the steady state \cite{challet00b, cavagna00},
which also reveals the physical meaning of $\Gamma$ as learning rate.

In the asymmetric phase,
the final state of the system and hence the volatility are independent of $\Gamma$ \cite{marsili01b}.
In the symmetric phase,
the final state of the system is dependent on $\Gamma$ and 
the volatility of the system increases with increasing $\Gamma$, 
provided that the system has reached the steady state \cite{cavagna00}.
This property of the game is in contrast to the ordinary physical systems,
where fluctuations increase with increasing temperature.
In the Minority Game,
fluctuations increase with increasing $\Gamma$,
i.e. decreasing temperature, 
as $\Gamma$ is implemented as individual inverse temperature in choosing strategies.
As a result,
in addition to inverse temperature or learning rate,
$\Gamma$ can also be interpreted as {\it collective} or {\it global effective temperature} 
of the whole system,
since global fluctuations increase with $\Gamma$.
In contrast to $\sigma^2$,
predictability $H$ is independent of $\Gamma$ in both symmetric and asymmetric phases \cite{marsili01b}.

In addition to dependence on $\Gamma$ in the symmetric phase,
the final state of the system is dependence on the initial conditions.
For games with the same set of strategies among agents (identical quenched disorders),
the final state of the system is dependent on the bias of initial virtual scores 
(heterogeneous initial conditions of annealed variables) of the strategies \cite{marsili01b, wong05}.
For the case of $S=2$,
the volatility of the system is smaller if larger differences are assigned to the initial virtual scores (i.e. initial bias)
of the two strategies,
given the same $\Gamma$ is implemented \cite{marsili00}.
A system with final condition depending on the initial state of annealed variable 
corresponds to the spin glass phase,
or the replica symmetry breaking (RSB) phenomenon in physical systems.
Thus,
the symmetric phase also corresponds to the behaviours of broken replica symmetry.
On the other hand,
in the asymmetric phase,
the final state of the system and hence the values of volatility 
is independent of the initial conditions.
Thus,
the asymmetric phase also corresponds to the replica symmetry (RS)
phase in physical systems.

\subsection{The cases of $S>2$}

\begin{figure}
\centerline{\epsfig{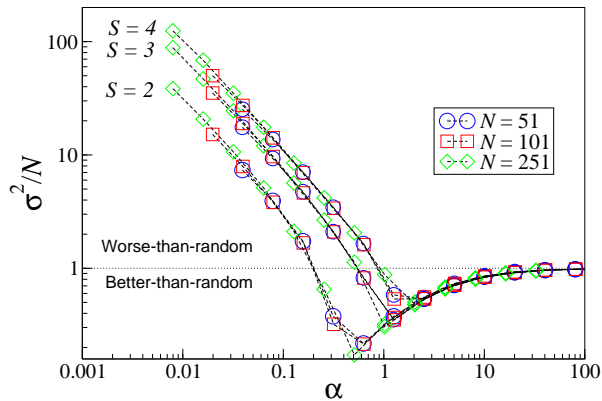}}
\caption{
The simulation results of the volatility as a function of the control parameter $\alpha=2^M/N$
for games with $S=2, 3, 4$ strategies for each agent averaged over 100 samples.
Endogenous information and linear payoff are adopted in these simulations.
Volatility generally increases with the number of strategies $S$ per agent.
Data collapse of volatility is shown for different values of $S$.
}
\label{multiS}
\end{figure}

Finally,
the behaviours of the game are also dependent on $S$,
the number of strategies that each agent hold.
As shown in Fig.~(\ref{multiS}) where the volatility is plotted against $\alpha=2^M/N$,
the volatility of the system is dependent on $S$.
Data collapse of volatility with different values of $N$ and $M$ is still shown by 
plotting volatility against $\alpha$,
for each value of $S$.
While the generic shape of the curves preserve when $S$ increases,
the points of minimum volatility shift to the right which suggests that 
the phase transition point is a function of $S$.
It is also suggested in \cite{challet98, zhang98} that for the cases of $S>2$,
instead of $\alpha=2^M/N$,
the important control parameter should be $2^{M+1}/SN$.
Since $2^{M+1}$ is the number of important strategies in the reduced strategy space
and $SN$ is the total number of strategies held by all agents,
when $2^{M+1}<SN$,
some agents are using identical strategies and crowds and anticrowds are formed.
On the other hand,
when $2^{M+1}>SN$,
most agents are using independent strategies and crowds are not formed.
Numerical solutions from replica approach for different values of $S$
show the relation of $\alpha_c(S)\approx\alpha_c(S=2)+(S-2)/2$
to a high degree of accuracy \cite{marsili00}.

\section{Variants of the Minority Game}
\label{secVariants}

After the first publication of the Arthur's bar problem and the Minority Game,
many variants of the game are established and studied by the physics community 
and also some economists.
Some of this variant models are developed to further simplify the Minority Game or
to include more features from the financial markets.
Most of the modifications include the use of different kind of strategies
and payoff functions,
the presence of different kinds of agents and the increased flexibility in 
participation of agents,
evolution of agents, replacement of poorly-performed agents by new agents and 
the individual concerns for capital.
In this section,
some of these variants are briefly introduced together with their physical significance
to the development of the Minority Game.
We leave their implications to the financial markets to Section \ref{secFin}.

\subsection{The Evolutionary Minority Game or the Genetic Model}
In 1999,
N. F. Johnson {\it et al.} \cite{johnson99} introduced the Evolutionary Minority Game which 
is usually quoted as EMG in literatures or the Genetic Model in later literatures.
From the name of EMG,
evolution of agents is an important feature added to the game.
In addition,
the strategies employed by agents are also major modifications.
Unlike the basic Minority Game,
all agents in EMG hold only one strategy $S=1$ and the strategy table
is identical for everyone.
For example in the case of $M=3$,
all agents hold one strategy as in Table \ref{strTable}.
Instead of having a column of fixed predictions,
this column record the most recent past winning action or choice
for the corresponding history.
Thus,
this strategy table is time dependent.
To make decisions,
all agents are assigned a different probability $p_i$ at the beginning,
with $0\leq p_i\leq 1$,
which is defined as the probability that agent $i$ acts according to the strategy table,
i.e. follow the recent winning action or the last outcome for that $M$-bit history.
With a probability $1-p_i$,
agent $i$ choose the choice opposite to the past winning action for that history.
This probability $p_i$ (rather than the strategy table) acts as a role of strategy in making decisions for agents
and is called ``strategy" in EMG or the ``gene" value in Genetic Model.
Hence,
the payoff or scores are rewarded or penalized subject to $p_i$.

To enhance the evolutionary property of the game,
agents are allowed to modify the $p_i$ if the scores fall below a threshold 
denoted by $d$,
where $d<0$,
which is sometimes known as the death score.
The new $p_i$ is being drawn with an equal probability in the range $(p_i-r/2, p_i+r/2)$ of width $r$,
with either periodic or reflective boundary condition at $p_i=0$ or $p_i=1$.
This corresponds to an evolution of strategy (the probability $p_i$), 
or the mutation of the gene value $p_i$ with mutation range $r$,
as a result the EMG is also known as the Genetic Model.
It is found that in the ordinary EMG with winning rule to be the minority group ($A(t)<N/2$),
the memory $M$ of the strategy table 
is not relevant in affecting the major features (include the steady state distribution ${\rm P}(p_i)$)
of the system \cite{burgos00},
but may be relevant with other winning rules (winning level other than $N/2$) \cite{kay04}.
Thus,
the volatility is independent of $M$ in the ordinary EMG,
in contrast to the basic Minority Game.

\begin{figure}
\centerline{\epsfig{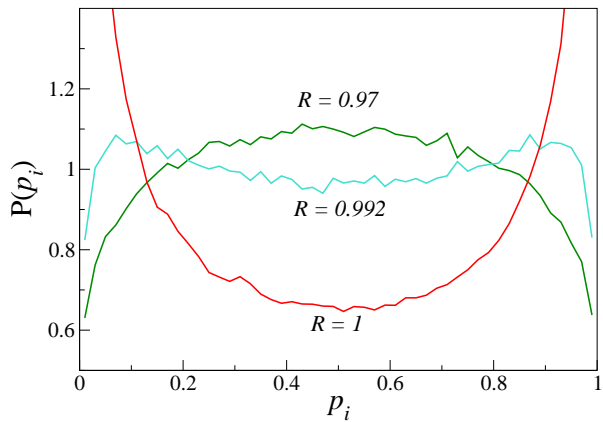}}
\caption{
The simulation results of the distribution ${\rm P}(p_i)$ in a game
of $N=10001$ agents with $d=-4$ and $M=3$.
The distribution is obtained after 120000 time steps and averaged over 50 simulations.
Self-segregation and clustering of agents at different values of $R$ are shown.
}
\label{pDis}
\end{figure}

In this ordinary model, 
one point is added to or deducted from the strategy $p_i$ for winning or losing predictions.
The possibilities of having a non-unity price-to-fine ratio $R$ are introduced in Hod {\it et al.} \cite{hod02, burgos00}.
For $R>1$,
agents are in a wealthy regime since the gain is larger than the lose in one game.
For $R<1$,
agents are in a tough regime.
The system behaviours are dependent on $R$, 
with two thresholds $R_c^{(1)}$ and $R_c^{(2)}$,
both $R_c^{(1)}$ and $R_c^{(2)}$ are less than and close to 1,
with $R_c^{(1)} > R_c^{(2)}$.
In the regime where $R>R_c^{(1)}$,
after a sufficient long time of evolution,
the agents self-segregated into two opposing group at $p_i=0,1$ 
and a "U-shaped" $P(p)$ distribution is found as shown in the case of $R=1$ in Fig. \ref{pDis}.
It implies agents tend to behave in an extreme way in a rich economy.
In the regime where $R<R_c^{(2)}$,
the agents cluster at $p_i=0.5$ and an "inverse-U-shaped" distribution is found as shown in the case of $R=0.97$ in Fig. \ref{pDis},
implying agents tend to be more cautious in a poor economy.

\subsection{The Thermal Minority Game}
In addition to the stochasticity in choosing strategies as introduced by Eq.(\ref{strProb})
discussed in the Section \ref{secPhys},
there are several other modifications from the basic Minority Game in the TMG \cite{cavagna99}.
In the original formulation,
the strategy is a vector in the $P$-dimensional real space ${\mathbf R}^P$ denoted by $\vec{a}_{i,s}$,
with $||\vec{a}_{i,s}||=\sqrt{P}$.
Thus,
the strategy space is the surface of the $P$-dimensional hypersphere
and the components of the strategies are continuous.
It is different from the discrete strategies in the basic Minority Game.
Every agent in the game draw $S$ vector strategies before the game starts.

The information processed by the strategies is a random vector $\vec{\eta}(t)$,
with a unit-length in ${\mathbf R}^P$.
The response or the bid of the strategy is no longer integer and is given
by the inner product of the strategies and the information,
i.e. $\vec{a}_{i,s}\cdot \vec{\eta}(t)$.
Hence, 
the attendance $A(t)$ is given by 
\begin{eqnarray}
\label{TMGattendance}
	A(t)=\sum_{i=1}^N \vec{a}_{i,s_i(t)}\cdot \vec{\eta}(t)
\end{eqnarray}
with $s_i(t)$ denotes the chosen strategy of agent $i$ at time $t$.
The cumulated payoff of strategy can be updated by
\begin{eqnarray}
\label{TMGpayoff}
	U_{i,s}(t+1)=U_{i,s}(t)-A(t)[\vec{a}_{i,s}\cdot \vec{\eta}(t)].
\end{eqnarray}
TMG can be considered as a continuous formulation of the Minority Game,
in which the game is no longer discrete and binary.
Since the response of the strategy in TMG is defined as the inner product,
i.e. a sum over all the $P$ entries of the vectors,
all components of the strategy have to predict at each round.
This is a difference from the basic Minority Game where at each round,
only one of the $P$ predictions on the strategy is effective.
Despite these differences and the continuous formulation,
TMG reproduces the same collective behaviours as the basic MG \cite{cavagna99, challet00b}.

\subsection{The simple Minority Game without information}
In this simplified Minority Game,
no information are given to agents and thus they have no strategy tables.
Agents choose between the choice $+1$ and $-1$ according to the following probabilities,
\begin{eqnarray}
\label{simMGprob}
	{\rm Prob}\{a_i(t)=\pm 1\} \equiv \frac{e^{\pm U_i(t)}}{ e^{U_i(t)} + e^{-U_i(t)} }
\end{eqnarray}
and $U_i(t)$ is updated by 
\begin{eqnarray}
\label{simMGU}
	U_i(t+1) = U_i(t) - \frac{\Gamma}{N}A(t)
\end{eqnarray}
where $U_i(t)$ can be considered as the virtual score for agent $i$ to 
make a decision of ``+1",
and $-U_i(t)$ correspondingly the virtual score for agent $i$ 
to make a decision of ``-1".
If $U_i(t)>0$,
the past experience of the agents shows that it is more successful to take action $a_i(t)=+1$,
and vice versa.
The learning rate or the temperature $\Gamma$ is implemented in this model.
This model gives a very simple analytic explanation on the system's dependence 
on $\Gamma$.
For $\Gamma<\Gamma_c$,
the volatility is found to be proportional to $N$, 
i.e. $\sigma^2\propto N$.
For $\Gamma>\Gamma_c$,
$\sigma^2\propto N^2$.
$\Gamma_c$ is found to be dependent on the initial conditions $U_i(0)$.
In addition,
$\sigma^2$ decreases with increasing $U_i(0)$.
Similar dependence of the volatility on $\Gamma$ and initial conditions,
and the dependence of $\Gamma_c$ on initial conditions are also found in the basic Minority Game.

\subsection{The Grand-canonical Minority Game}
The grand-canonical Minority Game (abbreviated, GCMG) refers to a subclass 
of Minority Games where the number of agents who actively participate in the 
market is variable.
Instead of summing over all $N$ agents,
the collective action or the attendance $A(t)$ is effectively the sum of the actions of active agents at time $t$.
Agents can be active or inactive at any time,
depending on their potential profitability from the market.
The term ``grand-canonical" origin from the grand-canonical ensemble in statistical mechanics
where the number of particles in the observing system is variable.
In most of the formulations \cite{slanina99, challet03, giardina03},
when the highest virtual score of the strategies that an agent hold is below some
threshold or $\epsilon t$ 
(where $\epsilon$ is usually a positive constant being referred to as the {\it interest rate} and 
$t$ is the number of rounds or time steps proceeded from the beginning of the game),
the agent refrain from participating in that round of the game.
It is equivalent to the addition of an {\it inactive strategy} for every agent from which agents become inactive,
and the virtual score of this strategy is $\epsilon t$.
Physically,
it corresponds to circumstances of gaining an interest of $\epsilon$ at each time step 
by keeping the capital in the form of cash,
so agents would only participate in the market if the gain from investments in the market
is greater than the interest rate.
In some other formulations,
instead of the virtual score of the strategies,
the real score of the agents is used to compare with the interest rate \cite{jefferies01}.
Winning probabilities of strategies within a certain time horizon are also considered \cite{lamper01, jefferies01}.
Individual capital concerns can also be implemented to achieve the grand-canonical 
nature of the game \cite{challet00c, giardina03},
in which agents vary their investment size at each time step by considering 
risk, 
gain potential or their limited capital.

These grand-canonical modifications from the Minority Game are considered to be 
important and crucial in producing the stylized facts of financial markets in the Minority Game models 
\cite{jefferies01, challet00c, challet03, challet01, giardina03},
while preserving the two-phase structure of the predictable and unpredictable phase.
The stylized facts being reproduced in the models include the fat-tail volatility or price return distributions
and volatility clustering,
when the systems are close to the critical state.
Numerical tests and analytical attempts are carried out in the critical regime of the models.
These models serve as a tool for physicists in understanding how macroscopic features are produced
from the microscopic dynamics of individuals,
which also support the conjecture of self-organized criticality of the financial market 
in which the financial market is always close to or attracted to the critical state.

\section{Analytic approaches on the Minority Game}
\label{secAnalytic}

There are several analytic approaches in solving the Minority Game.
Most of the approaches are based on the models of the basic Minority Game
with little modifications or simplifications.
It is found that in the asymmetric phase with $\alpha>\alpha_c$,
both equilibrium approaches and dynamics approaches are likely to describe the same behaviours 
of the system,
and the equilibrium approach based on the minimization of $H$ give an analytic solution in this phase.
\cite{challet00d, challet00, marsili00, marsili01b, garrahan00}.
For the symmetric phase with $\alpha<\alpha_c$,
fluctuations in the dynamics has to be considered and the solution is dependent on initial conditions. 
The final state of the system is sensitive to initial conditions and perturbations in the dynamics
\cite{heimel01, marsili01b, garrahan00}.
In this case,
solution is available in the limit of $\Gamma\rightarrow 0$ or asymptotic behaviours 
can be obtained in the limit of $\alpha\rightarrow 0$.

One of the early approach in solving the Minority Game is the crowd-anticrowd theory
which provide a qualitative explanation of the volatility dependence on brain size $M$ \cite{johnson99b, hart00, hart01}.
Consider the reduced strategy space (RSS) with strategies $R=1\cdots 2^M$ to be the uncorrelated
strategies, 
$\bar{R}$ to be the anti-correlated strategy of $R$ (i.e. $R$ and $\bar{R}$ always have opposite decisions),
which constituent the $2^{M+1}$ strategies in the RSS.
We denote $n_R$ to be the number of agent using the strategy $R$,
$n_{\bar{R}}$ to be the number of agents using $\bar{R}$ and $\langle\cdots\rangle$ to be time averaging.
For $R$ and $R'$ to be uncorrelated strategies,
the time average $\langle \sum_{R\neq R'}(n_R-n_{\bar{R}})(n_{R'}-n_{\bar{R'}})\rangle=0$
and thus the volatility $\sigma^2$ can be expressed as
\begin{eqnarray}
\label{crowdanitcrowd}
	\sigma^2=\sum_{R=1}^{2^M} (n_R-n_{\bar{R}})^2
\end{eqnarray}
which physically corresponds to the contribution to the global volatility from
each crowd-anticrowd pair ($R$, $\bar{R}$),
as $R$ and $\bar{R}$ are always making opposite decisions.
If we consider a uniform distribution of all strategy combination among agents
at the beginning of the game,
$n_R$ and $n_{\bar{R}}$ can be determined from the ranking of virtual scores of strategies,
since agents are always using the best strategies they hold.
In this case,
the strategy with highest virtual point would be the most popular strategy,
while its anti-correlated partner would have lowest virtual point and becomes the least popular strategy.
This happen for small $M$ where the number of strategies is small
and a large number of agents are using the best strategy.
On the contrary
only a small number of agents are using its anti-correlated strategy,
leading to a large $|n_R-n_{\bar{R}}|$ and a large volatility.
For the cases of large $M$,
even for the best strategy,
$n_R$ is relatively small and the $n_{\bar{R}}$ may have a similar magnitude as $n_R$,
leading to a small $|n_R-n_{\bar{R}}|$ and a small volatility.
This qualitatively explains the behaviours of volatility with $\alpha$ based on the size of crowd-anticrowd pairs.
The presence of temperature is also considered in the extended crowd-anticrowd approach \cite{hart00}.

In addition to this crowd-anticrowd theory,
full analytic approach can be developed.
To solve the Minority Game analytically,
we employ some convenient notation change which make the tools in statistical physics 
more applicable \cite{challet00d, challet00, marsili00}.
For the case of $S=2$,
we denote the first strategy of an agent to be ``+1" while the second one to be ``-1" whereas
the best strategy of agent $i$ at time $t$ is now expressed as $s_i(t)=\pm 1$.
The real bid $a_i(t)$ of agent $i$ at time $t$ can then be expressed as 
\begin{eqnarray}
\label{spin}
	a_i(t)=a_{i,s_i(t)}^{\mu(t)}=\omega_i^{\mu(t)}+s_i(t)\xi_i^{\mu(t)}
\end{eqnarray}
where $\omega_i^\mu=(a_{i,+}^\mu+a_{i,-}^\mu)/2$ and $\xi_i^\mu=(a_{i,+}^\mu-a_{i,-}^\mu)/2$.
$\omega_i^\mu$ and $\xi_i^\mu$ are quenched disorders and are fixed at the beginning of the game.
$\omega_i^\mu, \xi_i^\mu=0, \pm 1$ and $\omega_i^\mu\xi_i^\mu=0$ for all $\mu$.
$s_i(t)$ is the dynamic variable and becomes explicit in the action of agents,
corresponding to the Ising spins in physical systems. 
Thus,
the attendance can be expressed as a function of spin $s_i(t)$ given by
\begin{eqnarray}
\label{attendanceTheory}
	A(t)=\Omega^{\mu(t)}+\sum_{i=1}^N\xi_i^{\mu(t)}s_i(t)
\end{eqnarray}
where $\Omega^{\mu}=\sum_i \omega_i^{\mu}$.

Other than the spin $s_i(t)$,
the virtual scores of the strategies are also dynamic and we denote the 
difference of the virtual scores of the two strategies of agent $i$ to be $Y_i(t)$ given by
\begin{eqnarray}
\label{eqY}
	Y_i(t)=\frac{\Gamma}{2}(U_{i,+}(t)-U_{i,-}(t)).
\end{eqnarray}
This $Y_i(t)$ determines the relative probabilities of using the 2 strategies with ``inverse temperature" $\Gamma$
and is updated by
\begin{eqnarray}
\label{updateY}
	Y_i(t+1)=Y_i(t)-\frac{\Gamma}{N}\xi_i^{\mu(t)}A(t)
\end{eqnarray}
which is given by the update of the individual virtual scores $U_{i,+}(t)$ and $U_{i,-}(t)$ in 
Eq. (\ref{payoff}) with a factor of $1/N$ in the last term.
Thus the probabilities Eq.~(\ref{strProb}) for using the strategies $s_i(t)=\pm 1$ at time $t$ becomes
\begin{eqnarray}
\label{strProbTanh}
	{\rm Prob}\{s_i(t) = \pm 1\} =\pi_{i,\pm} = \frac{1\pm \tanh Y_i(t)}{2}
\end{eqnarray}
From this equation, 
we can calculate the time average of $s_i(t)$ at equilibrium with probabilities Eq. (\ref{strProbTanh}),
denoted by $m_i$,
to be
\begin{eqnarray}
\label{eqm}
	m_i=\langle s_i\rangle=\langle\tanh(Y_i)\rangle
\end{eqnarray}
The system will be stationary with $\langle Y_i\rangle\sim v_i t$,
corresponding to a stationary state solution of the set of $m_i$.
From Eq. (\ref{updateY}), 
$v_i$ can be expressed as 
\begin{eqnarray}
\label{eqv}
	v_i=-\overline{\Omega\xi_i}-\sum_{j=1}^N\overline{\xi_i\xi_j}m_j
\end{eqnarray}
where $\overline{\cdots}$ denotes the average over $\mu$.
For $v_i\neq 0$, 
$\langle Y_i\rangle$ diverges to $\pm\infty$ and giving $m_i=\pm 1$,
corresponding to the frozen agents who always use the same strategy.
For $v_i=0$,
$\langle Y_i\rangle$ remains finite even after a long time and $|m_i|<1$,
corresponding to the fickle agents who always switch their active strategy
even in the stationary state of the game.
We can identify $\overline{\Omega\xi_i}+\sum_{j\neq i} \overline{\xi_i\xi_j} m_j$ to be an external field
while $\overline{\xi_i^2}$ to be the self-interaction of agent $i$.
For an agent to be frozen,
the magnitude of external field has to be greater than the self-interaction.
In order to have fickle agents in the stationary state,
the self-interaction term is crucial.

We note that the above equation of $v_i$ in Eq. (\ref{eqv}) and the corresponding conditions of frozen
and fickle agents are equivalent to the minimization of predictability $H$,
with $H$ written in the form of
\begin{eqnarray}
\label{theoryH}
	H=\frac{1}{P}\sum_{\mu=1}^P\big[\Omega^\mu+\sum_{i=1}^{N}\xi_i^\mu m_i\big]^2
\end{eqnarray}
Since $m_i$'s are bounded in the range $[-1,+1]$,
$H$ either attains its minimum at $d H/d m_i=0$,
giving $\overline{\Omega\xi_i}+\sum_j \overline{\xi_i\xi_j} m_j=0$ (fickle agents) or 
at the boundary of the range $[-1,+1]$ of $m_i$,
giving $m_i=\pm 1$ (frozen agents).
Thus,
we can identify $H$ as the Hamiltonian where
the stationary state of the system is the ground state which minimizes the Hamiltonian.
From Eq. (\ref{attendanceTheory}),
the volatility of the system can be expressed as
\begin{eqnarray}
\label{varianceTheory}
	\sigma^2=H+\sum_{i=1}^N\overline{\xi^2}(1-m_i^2)+\sum_{i\neq j}\overline{\xi_i\xi_j}\langle(\tanh Y_i-m_i)(\tanh Y_j-m_j)\rangle
\end{eqnarray}
The last term involves the fluctuations around the average behaviour of the agents
and is related to the dynamics of the system.

Identifying $H$ as Hamiltonian reduces the problem into a conventional physical problem of finding the ground 
state of the system by minimizing the Hamiltonian.
Solving the problem involves averaging the quantity $\ln Z$ over quenched disorders 
corresponding to the strategies $a_{i, \pm}^{\mu}$ (now represented by $\omega_i^{\mu} $ and $\xi_i^{\mu}$)
given to the agents at the beginning of the game.
We note that the system is a fully connected system in which agents 
interact with all other agents,
and can be handled by the {\it replica} approach as in spin glass models,
under the assumption of replica symmetry.
Given the fraction of frozen agents to be $\phi$,
which can be expressed as a function of $\alpha$,
$\alpha_c=0.3374\dots$ \cite{challet00} is found to be the solution of the equation
\begin{eqnarray}
\label{eqphi}
	\alpha=1-\phi(\alpha)
\end{eqnarray}
In addition,
this approach allows us to see that the system 
with different initial condition converges to the same unique solution,
corresponding to a single minima of $H$ in the phase $\alpha>\alpha_c$ 
(replica symmetry).
This method allows us to get a complete solution for the Minority Game for all $\Gamma$ 
in the phase $\alpha>\alpha_c$.
Macroscopic quantities such as $\sigma^2$ and $H$ can be analytically calculated.
For $\alpha<\alpha_c$, 
there are multiple minimas of $H=0$ and the system's final state is not unique  (replica symmetry breaking)
and depends on its initial state.
In this case, 
dynamics has to be considered.
Breaking of replica symmetry is also considered in the case with market impact \cite{deMartino01}.

We notice that the in the long run,
the characteristic time in the dynamics are approximately proportional to $N$,
where all agents observe the performance of their strategies among all $P$ states with $P=\alpha N$.
This characteristic time is also inversely proportional to $\Gamma$ since
the dynamics of scores take a time of approximately $1/\Gamma$ to adapt a change of scores,
as discussed before.
The real time $t$ can then be rescaled as
\begin{eqnarray}
\label{tau}
 	\tau=\frac{\Gamma}{N}t
\end{eqnarray}
in which one characteristic time step $\tau$ in the system corresponds to 
$N/\Gamma$ real time steps $t$.
This is the reason for the systems with small $\Gamma$ having a convergence time of 
$N/\Gamma$.
We can hence write down a dynamical equation for $Y_i$ in the rescaled time by denoting the 
variable $y_i(\tau)=Y_i(N\tau/\Gamma)$ which gives
\begin{eqnarray}
\label{eqCLT}
	\frac{d y_i}{d\tau}=-\overline{\Omega\xi_i}-\sum_{j=1}^N\overline{\xi_i\xi_j}\tanh(y_i)+\zeta_i
\end{eqnarray}
where the first two terms on the right hand side represents the average behavior of 
agents obtained by the average frequency they play their strategies \cite{marsili01b}.
These two terms are considered to be deterministic.
The last term $\zeta_i$ represents the noise or the fluctuations around the average behaviour.
The properties of these fluctuation is given by 
\begin{eqnarray}
\label{zeta}
	&&\langle\zeta(\tau)\rangle=0	\\
\label{zeta2}
	&&\langle\zeta(\tau)\zeta(\tau')\rangle\cong\frac{\Gamma\sigma^2}{N}\overline{\xi_i\xi_j}\delta(\tau-\tau')
\end{eqnarray}
By writing down the Fokker-Planck equation for the probability distribution $P(\{y_i\},t)$ \cite{marsili01b},
much physical implications can be obtained.
We first note that the noise covariance Eq. (\ref{zeta2}) is linearly related to $\Gamma$,
revealing the role of $\Gamma$ as the global temperature of the system.
When $\Gamma\rightarrow 0$,
the noise covariance vanishes and the minimization of $H$ gives a valid solution,
even for $\alpha<\alpha_c$.
It can also be deduced that in the asymmetric phase,
the last term in Eq. (\ref{varianceTheory}) vanishes 
such that $\sigma^2$ is independent of $\Gamma$ and initial conditions.
In the symmetric phase,
this last term does not vanishes and $\sigma^2$ is dependent on both $\Gamma$ and initial conditions.

An alternative approach to derive dynamical equations is the generating functional approach \cite{heimel01, heimel01a, coolen01},
which monitors the dynamics using path integrals over time.
The approach was first used on the {\it batch} update version of the Minority Game,
in which agents update their virtual scores only after a batch of $P$ time steps
and with $\Gamma\rightarrow\infty$ as in the basic Minority Game.
The quenched disorder can be averaged out in the dynamical equations 
and in the limit of $N\rightarrow \infty$,
we obtain a representative ``single" agent dynamical equation with the variable $y(t)$,
where $y(t)$ represents the difference in the 
virtual scores of the two strategies of this ``single" agent after the $t$-th batch.
The dynamics is stochastic but non-Markovian in nature,
and can be extended to region inaccessible by the replica method.
This method again confirm the relation of Eq. (\ref{eqphi}) and giving the same value of $\alpha_c$ \cite{heimel01}
For $\alpha>\alpha_c$,
the fraction of frozen agents $\phi$ is obtained analytically and
the volatility is calculated to a high accuracy.
For $\alpha<\alpha_c$ in the limit of $\alpha\rightarrow 0$,
$\sigma$ is shown to diverge as $\sigma\sim\alpha^{-1/2}$ for $y(0)<y_c$, 
and vanishes as $\sigma\sim\alpha^{1/2}$ for $y(0)>y_c$,
with $y_c\approx 0.242$ \cite{heimel01}.
This approach was later extended to the case of on-line update (update of virtual score after every step)
and the cases of $\Gamma<\infty$ \cite{coolen01, heimel01a}

\section{Minority Games and Financial Markets}
\label{secFin}

The basic Minority Game model is a simple model which is used
to describe the possible interaction of investors in the financial markets.
Despite its simplicity,
some variants of the game show certain predictive abilities on real financial data 
\cite{jefferies01, lamper01, yeung08}.
Though Minority Games are simple,
they setup a framework of agent-based models from which
sophisticated trading models can be built,
and implementation on real trading may be possible.
Although these sophisticated models are usually used for private trading and may not be open to public,
Minority Games are still a useful tool to understand the dynamics in financial markets.
There are several fundamental difference between the basic game and the markets.
The basic Minority Game is a negative sum game in which the sum of gain of all agents is negative.
Agents do not concern about capital and cannot refrain from participation even if they
found the game unprofitable.
Whether the simple payoff function in Eq. (\ref{payoff}) correctly represents the 
evaluation of strategies by real investor is questionable.
We also note that the symmetric phase corresponds to a phase of information efficiency
in which the game becomes unpredictable.

Although the basic Minority Game gives a very colourful collective behaviours 
of agents from a simple interaction and dynamics,
some details can affect the behaviours 
and modifications have to be made in order to draw a more direct correspondence 
of Minority Games to real financial markets.
Some variants of the Minority Game are modified to study a particular 
issue or aspect of the real markets.
While with the introduction of several financial aspects,
some variants of the game lead to a more realistic model
of the market 
and at the same time complicating the models.
Among different aspects,
one of the primary issue is to draw analogy to trading
where price dynamics has to be introduced in the Minority Game.
A common price dynamics used in the game is to relate 
the attendance $A(t)$ to the price $p(t)$, 
and thus the return $r(t)$ in trading is 
given by 
\begin{eqnarray}
\label{price}
	r(t)\equiv\log[p(t+1)]-\log[p(t)]=\frac{A(t)}{\lambda}
\end{eqnarray}
where $\lambda$ is called the {\it liquidity} which is used to control the sensitivity 
of price on attendance.
With this or similar price dynamics,
trading process can be defined in the game.

After the introduction of price dynamics,
the issue of payoff function was also addressed.
The mixed minority-majority game originally proposed by M. Marsili \cite{marsili01}
is based on a a simplified version of the Minority Game in which there is no strategy table
and no information (as discussed in Section \ref{secVariants}).
The payoff function is this model is based on the expectations of the agents
on the price change in the next steps.
For simplicity, 
we consider the expectation $E_i[A(t+1)|t]$ of agent $i$ on the attendance $A(t+1)$ in the next step,
which is expressed as
\begin{eqnarray}
\label{expectation}
	E_i[A(t+1)|t]=-\Phi_i A(t)
\end{eqnarray}
For $\Phi_i>0$,
agents expect the attendance in the next step to be negatively correlated with that in the 
present step (i.e. price fluctuates),
revealing the minority nature of agents and they are called {\it fundamentalists}
or {\it contrarian} agents.
For $\Phi_i<0$,
agents expects the attendance in the next step to be positively correlated with that in the
present step (i.e. price trend develops),
revealing to the majority nature of the agents and they are called {\it trend followers}.
For both fundamentalists and trend followers,
if they expect the price to go up the next step,
buying is considered to be profitable,
and vice versa.
Thus, 
the payoff function $\delta U_i(t)=U_i(t+1)-U_i(t)$ is proportional to the product of the current decision $a_i(t)$
and the expectation of price change in the next step given by
\begin{eqnarray}
\label{minmajPayoff}
	\delta U_i(t)\propto a_i(t)E_i[A(t+1)|t]=-\Phi_i a_i(t)A(t)
\end{eqnarray}
with $\Phi_i>0$ and $\Phi_i<0$ corresponding to fundamentalists and trend followers respectively.
Hence,
fundamentalists are considered to be playing a minority game while
trend followers playing a majority game.

The two kinds of agents interact in the same game and
it was found that the ratio of fundamentalists to trend followers is important in affecting the 
behaviours of the system.
If more than half of the agents are fundamentalists,
the fundamentalists prevail and the game is minority in nature
with $\langle A(t+1)A(t)\rangle<0$.
On the other hand,
if more than half of the agents are trend followers,
the trend followers prevail and the game is majority in nature
with $\langle A(t+1)A(t)\rangle>0$.
Thus,
the behaviours of both minority and majority agents are found to be self-sustained,
depending on the relative population of the agents.

The $\$$-game \cite{andersen03} shares some similarity of the majority nature
of the trend followers in the mixed minority-majority game,
but with a crucial difference of using the real attendance of the next step,
not the expectations of agents,
in the payoff function.
In $\$$-game,
agents are again equipped with strategy tables and the virtual score of strategy $s$ 
is updated according to 
\begin{eqnarray}
\label{simMajU}
	U_{i,s}(t+1) = U_{i,s}(t) + a_{i,s}^\mu(t-1)A(t).
\end{eqnarray}
According to this payoff scheme,
the present actions $a_{i,s}^\mu(t)$ would only change the payoff at the next step at $t+1$.
Suppose an agent buys an asset,
he gains by selling the asset the next step if the price rises,
and vice versa.
This payoff scheme aims to model the mode of one-step speculating in realistic markets
though agents are not restricted to act oppositely in consecutive steps in the model.
Bubble-like behaviours are found in the model,
in which agents buy (sell) and push up (down) the price,
leading to positive evaluations of the buying actions such that agents are more likely to buy (sell) again.
This process continues and a persistent price trend is observed.
This persistent price trend is not observed in real markets,
and can be eliminated from the model if agents concern their limited capital, 
risk or maximum holding of assets \cite{andersen03, giardina03, challet08, yeung08}.
To summarize the different payoff schemes in the Minority game,
majority game and the $\$$-game,
Table~\ref{payoffTable} shows the payoff functions 
of the three games.

\begin{table}
\centering
\begin{tabular}{|l||c|}
\hline
& $\delta U_{i,s}(t)$ \\
 \hline \hline
 Minority Game & $-a_{i,s}^\mu(t)A(t)$ \\
 & \\
 Majority Game & $a_{i,s}^\mu(t)A(t)$ \\
 & \\
 $\$$-game & $a_{i,s}^\mu(t-1)A(t)$ \\
 \hline 
\end{tabular}
\caption{
The payoff functions of the Minority Game, the majority game
and the $\$$-game, 
with $\Phi_i$ set to $\pm 1$ in the mixed minority-majority game.
}
\label{payoffTable}
\end{table}

Other than the payoff functions,
we consider the stationary state of collective behaviours in the system.
The stationary state of the Minority Game is not a Nash Equilibrium.
There are extremely large number of Nash Equilibrium in the Minority Game \cite{marsili01, deMartino01} and 
one example is $(N+1)/2$ agents always make an action of $a_i(t)=+1$ while 
$(N-1)/2$ agents always make an action of $a_i(t)=-1$.
In this case,
$\sigma=1$ and no individual has incentives to change his action by himself alone 
(the majority group change if any of the losers moves).
This state is not stationary in the game.
The stationary state of the game is described by the minimization of predictability $H$,
but Nash Equilibriums are states of minimum volatility $\sigma^2$.
Physically,
instead of competing with the other $N-1$ players,
agents are interacting with the total attendance $A(t)$ which include also its own action.
By subtracting their own actions from the attendance,
their cumulated payoffs Eq. (\ref{simMGU}) in the simplified minority game becomes
\begin{eqnarray}
\label{Uimpact}
	U_i(t+1) = U_i(t) - \frac{\Gamma}{N}[A(t)-\eta a_i(t)]
\end{eqnarray}
where $\eta$ denotes the {\it market impact}.
The Minority Game corresponds to the case of $\eta=0$ in which Nash Equilibrium is not attained.
In this simplified model,
$\eta>0$ brings heterogeneity to behaviours of agents and Nash Equilibrium is attained \cite{marsili01}.

In some variants of the Minority Game,
the role of participants are studied.
It was suggested that symbiotic relation is present between two kinds of traders \cite{zhang99, challet00d, challet00c},
namely the {\it producers} and the {\it speculators}.
Producers are agents who always participate and trade with only one strategy.
They have a primary interest in trading in the market for business or other reasons.
Speculators are agents who speculate and have no interest in the intrinsic values of the assets traded.
They can refrain from participating in the market at any time they found it unprofitable.
This model correspond to one of the grand-canonical Minority Games.
In this model,
it was found that the gain of producers are always negative but their loses decrease with 
increasing number of speculators,
since speculators provide liquidity to producers and make the market more unpredictable.
On the other hand,
the gain of speculators generally increases with the number of producers,
since producers provide more information to speculators which make the market more predictable.
As a result,
producers and speculators are symbiotic.

In addition to studying the role of producers and speculators,
the grand-canonical Minority Game (GCMG) plays a crucial role in understanding
financial markets \cite{challet03, jefferies01, giardina03}.
By introducing the grand-canonical nature to the model,
agents can choose to refrain from participating in the market when
they found the game unprofitable.
While observing the market as non-trading outsiders,
they can participate in the market again once they found it profitable.
The predictable and unpredictable phases are usually preserved in this class of models,
where fat-tail distributions of price return are found around the phase transition point.
These can be fitted by a power laws.
While outside the critical region,
the fluctuation distributions becomes gaussian.
In addition,
volatility clustering,
where high volatility are likely to cluster in time,
is also found and can be fitted by power laws or other forms of function \cite{challet00c, giardina03, challet03, bouchaud00}.
It suggests that the dynamics of agents is correlated in time.

The observation of power laws in the model coincides with the observations of fat-tail price return distributions
and volatility clustering in
real markets in the high frequency range \cite{liu99}.
While gaussian fluctuations are not found in real markets,
these properties of the model provides an important implications or conjectures of financial markets being 
in the critical state.
It also suggests the possibility of self-organized critical system as 
an explanation of the behaviours of financial markets.
In addition to power laws,
rescaling of financial market data of different frequencies \cite{mantegna95} also provides preliminary evidence 
of the property of scale invariance in time in critical systems.
Although power laws in financial markets may have origins other than critical phenomenon \cite{gabaix03},
these conjectures provide a potential perspective in understanding the dynamics and behaviours or the markets.
If financial markets exist as a critical phenomenon,
their behaviours can be understood qualitatively from the underlying nature of interactions
in similar systems within the same universality class.
In this case,
microscopic details of the systems are not crucial in affecting the generic
behaviours for systems in the same class.
As financial markets are observed to be operated close to informationally efficiency,
correspondingly,
the grand-canonical Minority Games show stylized facts near the critical
point of phase transition to the unpredictable phase,
i.e. the phase of informationally efficiency.
In some versions of the grand-canonical minority game,
rarely large fluctuations resulting from a sudden participation 
of a larger number of speculators are also found which draw analogy to market crashes.

\section{Future Directions}
\label{secFutureDirection}

In view of the exciting physical pictures bring along with the grand-canonical 
minority games,
more analytic works can be developed in understanding the dynamics of the critical regime 
around which the fat-tail distributions and the volatility clustering are found.
Formation of power laws and anomalous fluctuations may be understood with
the analytic tools. 
Analytic approach on the grand-canonical minority game also provides 
more clues in proving or disproving the conjecture of the financial market being 
a self-organized critical phenomenon.

On the other hand,
simple modeling works which reveal more the dynamics of the financial market are still possible.
Development of other simple models which draw direct analogy to the financial 
markets,
together with analytic solution is crucial in understanding how the markets work.
Other than the grand-canonical games,
some variants of the basic Minority Game are still simple and worth solving analytically.
Although analytic solution may not be available for complicated models
which introduce more and more realistic aspects into the game,
comprehensive modeling based on the inductive nature of agent-based models 
provides us a new perspective in understanding the financial markets.

Other than modeling,
efforts may be put in implementing the Minority-Game kind of strategies in real trading.
Although the strategies from the basic Minority Game and its variants 
may not accurately describe the strategies for real trading,
the simplicity of the strategy does leave us a large freedom in expanding,
modifying and tuning with respect to profitability attaining in real trading.
Using Minority Games as framework,
sophisticated real trading system can be built which include
a comprehensive picture of trading mechanism.
Predictive capacity may also be obtained from this kinds of agent-based model
which goes beyond the standard economic assumption of deductive agents and market efficiency.
All these directions show a great potential over the 
conventional statistical tools in financial analysis.
These developments may be used for private trading and 
may not be accessible through public and academic literatures.
Some of the works \cite{jefferies01, yeung08, lamper01} have already shown potentials in this direction of application.


\section*{Books and Reviews}

Challet, D., Marsili, M. and Zhang, Y.-C. (2005)
Minority Games.
Oxford University Press, Oxford, UK.

\noindent Coolen, A. A. C. (2004)
The Mathematical Theory of Minority Games.
Oxford University Press, Oxford, UK.

\noindent Johnson, N. F., Jefferies, P., and Hui, P. M. (2003)
Financial Market Complexity.
Oxford University Press, Oxford, UK.

\noindent Minority Game's website:
http://www.unifr.ch/econophysics/minority/

\end{document}